\documentstyle[12pt,epsf,epsfig,wrapfig]{article}

\begin{document}
\begin{titlepage}
\begin{center}

{EUROPEAN ORGANIZATION FOR NUCLEAR RESEARCH} 
\vskip 0.5 cm
~~~~~~~~~~~~~~~~~~~~~~~~~~~~~~~~~~~~~~~~~~~~~~~~~~~~~~~~~~~~~~~~~~~~~~~~~~~~
~~~~~~~~~CERN PH-TH/2009-156\\
~~~~~~~~~~~~~~~~~~~~~~~~~~~~~~~~~~~~~~~~~~~~~~~~~~~~~~~~~~~~~~~~~~~~~~~~
17 August 2009
\vskip 2cm 

{\Large\bf Some Remarks on Methods of QCD Analysis of \\[2mm]
 Polarized DIS Data}

\end{center}
\vskip 1 cm
\begin{center}
{\bf Elliot Leader}\\
{\it Imperial College London\\ Prince Consort Road, London SW7
2BW, England }
\vskip 0.5cm
{\bf Aleksander V. Sidorov}\\
{\it Bogoliubov Theoretical Laboratory\\
Joint Institute for Nuclear Research, 141980 Dubna, Russia }
\vskip 0.5cm
{\bf Dimiter B. Stamenov \\
{\it CERN, Theory Division, CH 1211 Geneva 23, Switzerland, \\
Institute for Nuclear Research and Nuclear Energy\\
Bulgarian Academy of Sciences\\
Blvd. Tsarigradsko Chaussee 72, Sofia 1784, Bulgaria }}
\end{center}

\vskip 0.7cm
\begin{abstract}
\hskip -5mm

The results on polarized parton densities (PDFs) obtained using 
different methods 
of QCD analysis of the present polarized DIS data are discussed. 
Their dependence on the method used in the analysis, accounting 
or not for the kinematic and dynamic $1/Q^2$ corrections to 
spin structure function $g_1$, is demonstrated. It is pointed 
out that the precise data in the preasymptotic region require 
a more careful matching of the QCD predictions to the data in 
this region in order to determine the polarized PDFs correctly. 

\vskip 1.0cm PACS numbers: 13.60.Hb, 12.38.-t, 14.20.Dh

\end{abstract}

\end{titlepage}

\newpage
\setcounter{page}{1}

\section{Introduction}

Our present knowledge of the spin structure of the nucleon comes
mainly from polarized inclusive and semi-inclusive DIS experiments
at SLAC, CERN, DESY and JLab, polarized proton-proton collisions
at RHIC and polarized photoproduction experiments. One of the
important and best studied aspects of this knowledge is the
determination, in a QCD framework, of the longitudinal polarized
parton densities and their first moments, which are related to the
spins carried by the quarks and gluons in the nucleon. Different
methods of analysis have been used in these studies. The aim of
this paper is to discuss and clarify how the results on polarized
PDFs depend on the method used in the QCD analysis.

One of the peculiarities of polarized DIS is that more than a half
of the present data are at moderate $Q^2$ and $W^2~(Q^2 \sim
1-4~\rm GeV^2,~4~\rm GeV^2 < W^2 < 10~\rm GeV^2$), or in the
so-called {\it preasymptotic} region. So, in contrast to the
unpolarized case, this region cannot be excluded from the
analysis, and the role of the $1/Q^2$ terms ({\it kinematic} -
$\gamma^2$ factor, target mass corrections, and {\it dynamic} -
higher twist corrections to the spin structure function $g_1$) in
the determination of the polarized PDFs has to be investigated.
This makes the QCD analysis of the data much more complicated and
difficult than in the unpolarized case.

\section{QCD framework for inclusive polarized DIS}

\subsection{Which data to chose for QCD fits}

The best manner to determine the polarized PDFs is to perform a
QCD fit to the data on $g_1/F_1$, which can be obtained if both
the $A_{||}$ and $A_{\perp}$ asymmetries are measured. In some
cases only $A_{||}$ is measured. One can write (D is the
depolarization factor and $\gamma^2=4M^2x^2/Q^2$)
\begin{equation}
{A_{||}\over {D}} = A_1 + \eta A_2 = (1+\gamma^2){{g_1}\over
{F_1}} + (\eta - \gamma)A_2, \label{A||}
\end{equation}
from which one sees that the quantity $A_{||}/D(1+\gamma^2)$ is a
good approximation of $g_1/F_1$ because the second term in the
second relation of (\ref {A||}) can be neglected in the
preasymptotic region too - the asymmetry $A_2$ is bounded and in
fact small, and multiplied in addition by a small kinematic factor
($\eta \approx \gamma$). 

The data on the photon-nucleon asymmetry $A_1$ are not suitable
for the determination of PDFs because the structure function $g_2$
is not well known in QCD and the approximation
\begin{equation}
(A_1)^{theor}= g_1/F_1-\gamma^2g_2/F_1\approx (g_1/F_1)^{theor}
\label{A1theor}
\end{equation}
used by some of the groups is {\it not} reasonable in the {\it
preasymptotic} region because $\gamma^2$ cannot be neglected.

Bearing in mind the remarks above, let us discuss in more detail
how to confront correctly the theoretical predictions to the
available polarized inclusive DIS data:

~~i) First of all, one should include in the QCD fit of the world
data {\it all} $g_1/F_1$ data available. These are the CLAS(p, d),
JLab/Hall A(n), SLAC E143(p, d) and E155(p, d) data
\cite{g1F1data}.\footnote{Note that excepting the E155
Collaboration, the other Collaborations present data on $A_1$ too. 
The
corresponding values of $A_1$ and $g_1/F_1$ at the same $(x,~Q^2)$
are different, which means that the term $\gamma^2g_2/F_1$ cannot
be really neglected in the preasymptotic region and a fit to $A_1$ 
data instead of $(g_1/F_1)$, approximating $(A_1)^{theor}$ with
$(g_1/F_1)^{theor}$ is {\it not} correct.}

~ii) For the rest of experiments: EMC(p), SMC(p, d)
\cite{CERNdata} and COMPASS(d) \cite{COMPASS} at CERN, HERMES(p,
d) \cite{HERMES} at DESY and E142(n), E154(n) at SLAC
\cite{SLAC_A1}, only data on $A_1$ are presented. In the
experiments at CERN and DESY only the asymmetry $A_{||}$ is
measured. However, for different reasons the approximations $A_{||}/D
\approx A_1 \approx g_1/F_1$ for CERN data and $A_{||}/[D(1-\eta\gamma)]\approx A_1 \approx g_1/F_1$ for HERMES data 
are good ones. For the experiments  at CERN, the $\gamma^2$ factor is 
very small and the term
$\gamma^2g_2/F_1$ can be neglected, while for the HERMES data the
approximation $g_2=0$ is used and the effect of the non-zero value
of $g_2$ is included in the systematic uncertainty of
$A_1$.\footnote{Note that for the final inclusive HERMES $A_1$ data 
\cite{HERMES07} the approximation $g_2=0$ is not used and the
relation $A_1 \approx g_1/F_1$ does not hold. } In the SLAC
E142(n) and E154(n) experiments both $A_{||}$ and $A_{\perp}$ have
been measured and $g_1/F_1$ data could have been extracted, but
the collaborations present only data on $A_1$. Bearing in mind the
kinematic region (E154) and the precision (E142) of these data,
the approximation $A_1/(1+\gamma^2) \approx g_1/F_1$ in Eq. (\ref{A||})
 for them is reasonable.

To summarize, in the pure DIS region $A_1 \approx g_1/F_1$ and it
does not matter which data are be used in the QCD analysis. This
is not the case when precise data in the preasymtotic region have
to be used too. In that case one has to confront the QCD
predictions to the data more carefully in order to extract the
polarized PDFs correctly.

\subsection{Methods of QCD analysis}

In QCD, one can split $g_1$ and $F_1$ into leading (LT) and dynamical higher twist (HT) pieces
\begin{equation}
g_1 = (g_1)_{\rm LT, TMC} + (g_1)_{\rm HT},~~~~F_1 = (F_1)_{\rm LT, 
TMC} + (F_1)_{\rm HT}. \label{LTHT}
\end{equation}
In the LT pieces in Eq. (\ref{LTHT}) the calculable target mass corrections (TMC) 
are included 
\begin{eqnarray}
g_1(x,Q^2)_{\rm LT, TMC}&=&g_1(x,Q^2)_{\rm LT} + g_1(x,Q^2)_{\rm TMC},\\
\nonumber
F_1(x,Q^2)_{\rm LT, TMC}&=&F_1(x,Q^2)_{\rm LT} + F_1(x,Q^2)_{\rm TMC}.
\label{LT_TMC}
\end{eqnarray}
They are inverse powers of $Q^2$ kinematic corrections, which, however, 
effectively belong to the LT part of $g_1$. 
Then, approximately
\begin{equation}
{g_1\over F_1} \approx {(g_1)_{\rm LT}\over (F_1)_{\rm LT}}[1 +
{(g_1)_{\rm TMC+HT}\over (g_1)_{\rm LT}} - {(F_1)_{\rm TMC+HT}\over
(F_1)_{\rm LT}}]. \label{g1ovF1}
\end{equation}
Note that the LT pieces $(g_1)_{\rm LT}$ and $(F_1)_{\rm LT}$ are 
expressed in terms of the polarized and unpolarized PDFs, respectively. 
In what follows only 
the first terms in the TMC and HT expansions will be considered
\begin{eqnarray}
g_1(x,Q^2)_{\rm TMC}&=&M^2/Q^2g^{(1)}_1(x,Q^2)_{\rm TMC} + {\cal O}(M^4/Q^4),\\
\nonumber
F_1(x,Q^2)_{\rm TMC}&=&M^2/Q^2F^{(1)}_1(x,Q^2)_{\rm TMC} + {\cal O}(M^4/Q^4);
\label{g1F1_TMC}
\end{eqnarray}
\begin{equation}
g_1(x, Q^2)_{\rm HT}=h^{g_1}(x)/Q^2 + {\cal O}(\Lambda^4/Q^4),~~
2xF_1(x, Q^2)_{\rm HT}=h^{2xF_1}(x)/Q^2 + {\cal O}(\Lambda^4/Q^4).
\label{HT}
\end{equation}
The first terms of the HT pieces in Eq. (\ref{HT}) are shown in Fig. 1 
and as seen, they are definitely different from zero and cannot be 
neglected in the preasymptotic region.
\begin{figure}[h]
\begin{center}
\mbox{\epsfig{figure=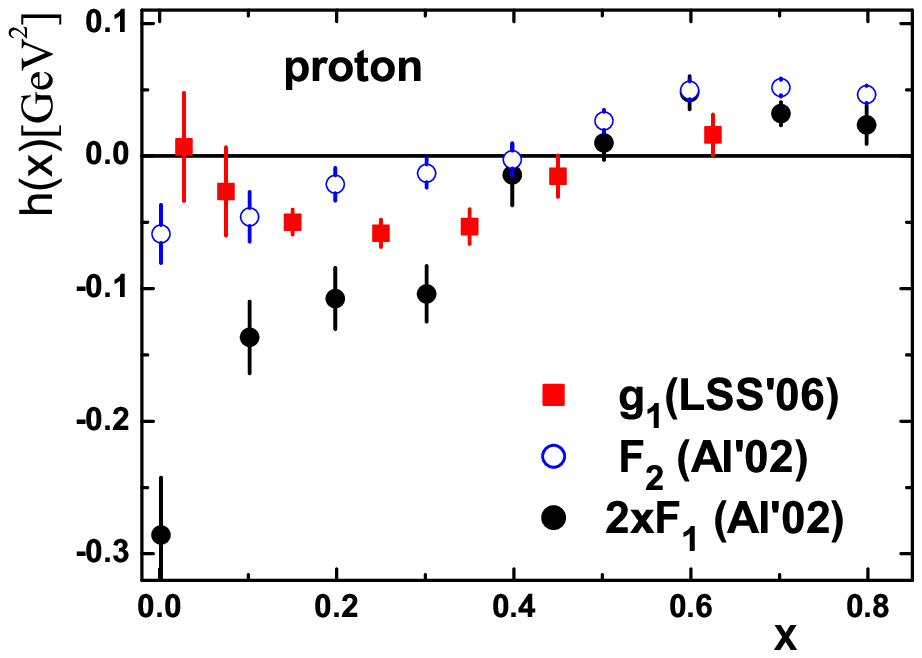,width=7.5cm,height=5.5cm}}
\end{center}
\centerline { {\small{\bf Figure 1.} HT corrections to $g_1$ \cite{LSS06},~$F_2$ and $2xF_1$ \cite{Alekhin} structure functions.  }}
\end{figure}

There are essentially two methods to fit the data - taking or NOT
taking into account the HT corrections to $g_1$. According to the
first \cite{LSS_HT}, the data on $g_1/F_1$ have been fitted
including the contribution of the first term $h(x)/Q^2$ in
$(g_1)_{\rm HT}$ and using the experimental data for the
unpolarized structure function $F_1$
\begin{equation}
\left[{g_1(x, Q^2)\over F_1(x, Q^2)}\right]_{exp}~\Leftrightarrow~
{{g_1(x,Q^2)_{\rm LT, TMC}+h^{g_1}(x)/Q^2}\over F_1(x,Q^2)_{exp}}~~~~(\rm
Method~I). \label{g1HTf1}
\end{equation}
According to the second approach \cite{GRSV} only the LT terms for
$g_1$ and $F_1$ in (\ref{LTHT}) have been used in the fit to the
$g_1/F_1$ data
\begin{equation}
\left[{g_1(x, Q^2)\over F_1(x, Q^2)}\right]_{exp}~\Leftrightarrow~
{{g_1(x,Q^2)_{\rm LT}}\over F_1(x,Q^2)_{\rm LT}}~~~~(\rm
Method~II). \label{g1f1_LT}
\end{equation}

It is obvious that the two methods are equivalent in the pure DIS
region where HT can be ignored. To be equivalent in the {\it
preasymptotic} region requires a cancellation between the ratios
$(g_1)_{\rm TMC+HT}/(g_1)_{\rm LT}$ and $(F_1)_{\rm TMC+HT}/(F_1)_{\rm
LT}$ in (\ref{g1ovF1}). Then $(g_1)_{\rm LT}$ obtained from the best 
fit to the data will coincide within the errors independently of the 
method which has been used. In Fig. 2 these ratios based on our  
results on target mass \cite{TMC} and higher twist \cite{LSS06} 
corrections to $g_1$, and 
the results on the unpolarized structure function $F_1$ are presented. 
Note that for the neutron target, 
$(g_1)_{\rm TMC+HT}$ is compared with $(g_1)_{\rm LT}{(F_1)_{\rm TMC+HT}\over (F_1)_{\rm LT}}$ because of a node-type behaviour of 
$(g_1)_{\rm LT}$. Also, LT means the NLO 
QCD approximation for both, $g_1$ and $F_1$. As seen from Fig. 2, 
(TMC+HT) corrections to $g_1$ and $F_1$ in the ratio $g_1/F_1$ do not 
cancel and ignoring them using the second method is incorrect and will impact on the determination of the polarized PDFs.\footnote{Note that
this result differs from our previous observation that the ratios
$(g_1)_{\rm HT}/(g_1)_{\rm LT}$ and $(F_1)_{\rm HT}/(F_1)_{\rm
LT}$ for a proton target approximately cancel for $x>0.15$ at $Q^2=2.5~GeV^2$ \cite{HTcancel}. However, in a more precise analysis one should account for the TMC corrections in Eq. (\ref{g1ovF1}). 
Note that previously, for the calculation of the ratio $(F_1)_{\rm HT}/(F_1)_{\rm LT}$ the results of \cite{Alekhin} were used, where a QCD analysis of the world unpolarized DIS data was performed using the cut $Q^2\geq 2.5~GeV^2$. This cut excludes a lot of data in the preasymptotic region which influence the determination of the HT corrections to $F_1$. Fig. 2 is based on a new analysis of the ratio $(F_1)_{\rm TMC+HT}/(F_1)_{\rm LT}$ \cite{Stamenov}. }
\begin{figure}[h]
\begin{center}
\begin{tabular}{cc}
\mbox{\epsfig{figure=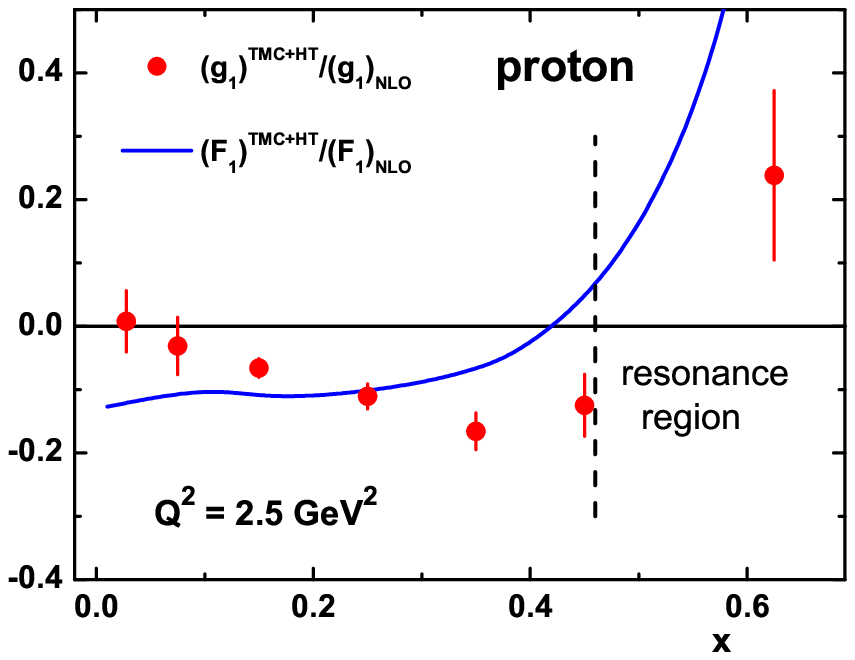,width=6.5cm,height=5cm}}&
\mbox{\epsfig{figure=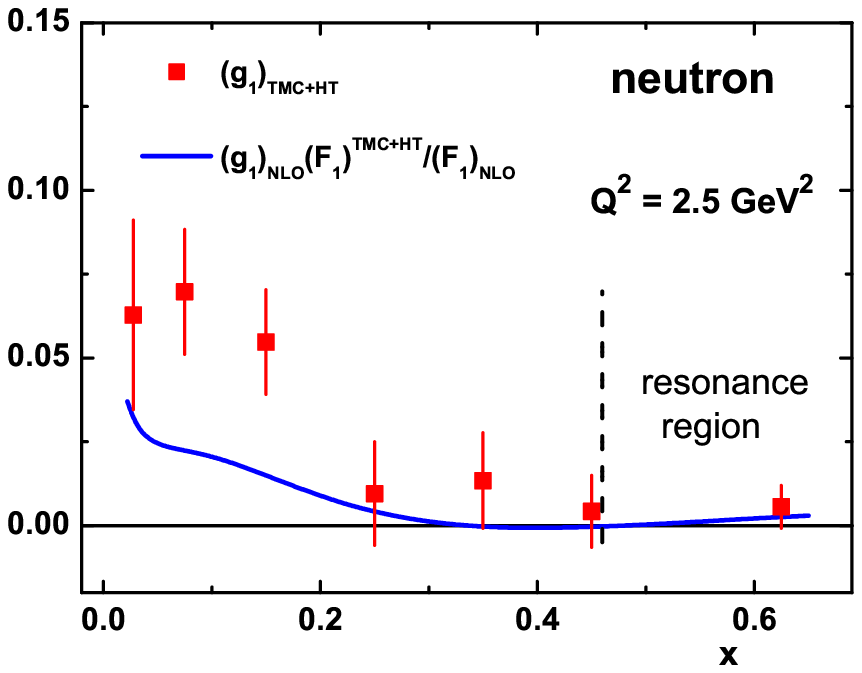,width=6.5cm,height=5.2cm}}\\
\end{tabular}
\end{center}
\small{\bf Figure 2.} Comparison of the ratios of (TMC+HT)/LT
for $g_1$ and $F_1$ structure functions for proton and neutron 
targets (see the text). 
\end{figure}

Modifications of the second method of analysis in which $F_1$ is
treated in different way are also presented in the literature:

~i) Blumlein, Bottcher \cite{BB} and COMPASS \cite{COMPASS}, where
instead of $(F_1)_{\rm LT}$ in (\ref{g1f1_LT}), $(F_1)_{exp}$ has
been used in the fit to the data
\begin{equation}
\left[{g_1(x, Q^2)\over F_1(x, Q^2)}\right]_{exp}~\Leftrightarrow~
{{g_1(x,Q^2)_{\rm LT}}\over F_1(x,Q^2)_{exp}}~; \label{g1f1_exp}
\end{equation}

ii) AAC Collaboration \cite{AAC}, where $F_1$ is expressed in terms 
of $F_2$ and $R$, and for them: $(F_2)_{\rm LT}$ and $R_{exp}$ have 
been used, respectively
\begin{equation}
A_1(x,Q^2)_{exp} \approx \left[{g_1(x, Q^2)\over F_1(x, Q^2)}\right]_{exp}~\Leftrightarrow~
{{g_1(x,Q^2)_{\rm LT}}\over {F_2(x,Q^2)_{\rm LT}}}
2x( 1+R(x, Q^2)_{exp})~. \label{AAC}
\end{equation}
As mentioned above, the approximation for $A_1$ in (\ref{AAC}) is not
correct for most of the data sets used in the fit. 

Note that when the second method or its modifications
are used, the HT effects of $g_1$ are apparently absorbed into the 
extracted PDFs, which thus differ from those determined in the 
presence of HT (for more details see the discussion below), but, of 
course, not all the data can be fitted satisfactorily. On other hand, the
extracted PDFs in the framework of the second method and
corresponding to fits (\ref{g1f1_LT} - \ref{AAC}) should all
be different due to the HT corrections to $F_1$ (see Fig. 1)
which are included in Eq. (\ref{g1f1_exp}) but not in Eq.
(\ref{g1f1_LT}), and only partly in Eq. (\ref{AAC}). So, using the different denominators in (\ref{g1f1_LT} -
\ref{AAC}) one will obtain different values for the free
parameters associated with the input polarized PDFs after fitting the data.

\section{Polarized Parton Densities}

We will discuss in this section in more detail how the results on
polarized PDFs depend on the method used for their determination.
To illustrate this dependence we will compare the NLO LSS'06 set
of polarized parton densities \cite {LSS06} determined by Method I
with those obtained by COMPASS \cite{COMPASS}, DSSV \cite{DSSV} and
AAC Collaboration \cite{AAC08} using Method II or its modifications.

\subsection{Comparison between LSS'06 and COMPASS PDFs}

To obtain the LSS'06 PDFs we used Method I (Eq. \ref{g1HTf1}) in
the QCD analysis of the the world data on polarized inclusive DIS
(\cite{g1F1data}-\cite{SLAC_A1}), i.e. the HT corrections to $g_1$
were taken into account. For the LT term we have used the NLO
QCD approximation in the ${\rm \overline{MS}}$ renormalization
scheme. In their analysis COMPASS has used Eq. (\ref{g1f1_exp}), but
the CLAS data from \cite{g1F1data} were not included in their
fit.\footnote{Using this method one cannot achieve an acceptable
value of $\chi^2$ for the CLAS data which are entirely
in the preasymptotic region \cite{Winmolders}.} In both the
analyses the experimental data for the unpolarized structure
function $F_1$ were used. Thus what is fitted by
$(g^N_1)_{\rm LT}(\rm COMPASS)$ is significantly different from
what is fitted by our $(g^N_1)_{\rm LT}(\rm LSS)$, i.e.
\begin{equation}
g^N_1(x, Q^2)_{\rm LT}({\rm COMPASS}) = g^N_1(x, Q^2)_{\rm
LT, TMC}({\rm LSS})+ h^N(x)/Q^2 \equiv (g_1^N)_{\rm tot}^{\rm LSS}~~~
(N=p,n,d). \label{LT_LSS_COMPASS}
\end{equation}
Note that for $(g_1)_{\rm LT}$ COMPASS has also used the NLO QCD
approximation in the ${\rm \overline{MS}}$ scheme. As a result,
$(g_1)_{\rm tot}^{\rm LSS}$ and $(g_1)_{\rm LT}({\rm COMPASS})$
obtained from the fit are almost identical, but the LT terms of
$g_1$ corresponding to LSS and COMPASS fits are different for $x <
0.1$ where the HT corrections to $g_1^d$ cannot be neglected. This
fact is illustrated in Fig. 3 for $g_1^d$ where the COMPASS data
are also presented. 
\begin{figure}[h]
\begin{center}
\mbox{\epsfig{figure=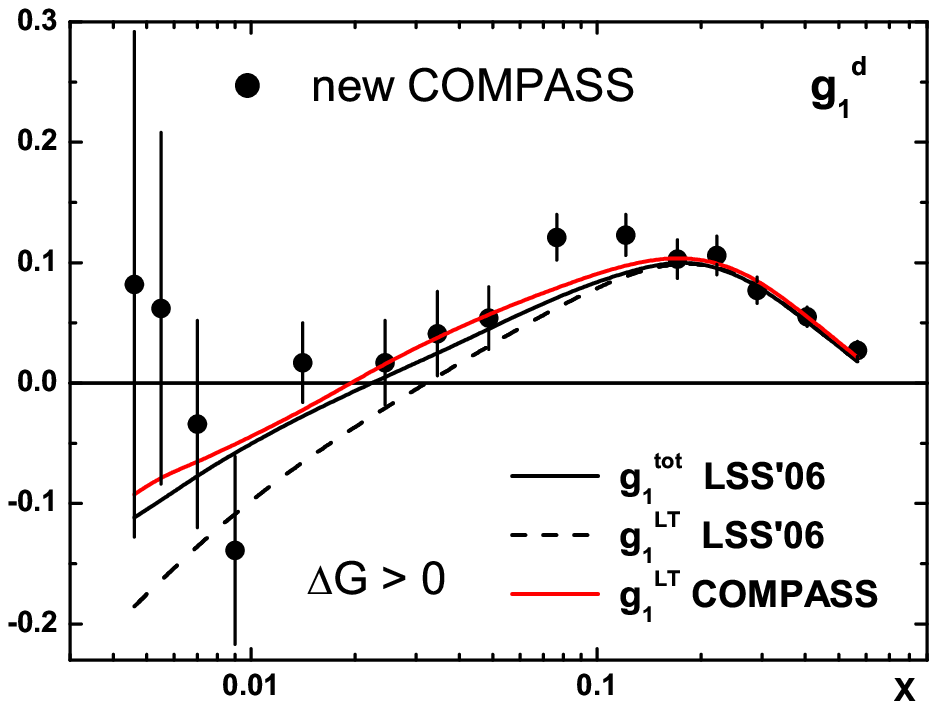,width=6.5cm,height=5.5cm}}
\end{center}
{\small {\bf Figure 3.} Comparison between $(g_1)_{tot}(LSS)$ and
$(g_1)_{LT}(COMPASS)$ obtained from the fit with the COMPASS data
at measured $x$ and $Q^2$. Error bars represent the total
(statistical and
 systematic) errors. The $(g_1)_{LT}(LSS)$ curve is also shown. }
\end{figure}
We have found that the HT contribution to
$(g^d_1)_{\rm tot}$, $h^d(x)/Q^2$, is positive and large, up to
40\% of the magnitude of $(g^d_1)_{\rm LT}$ in the small $x$
region, where $Q^2$ is small ($Q^2 \sim 1-3~GeV^2$). As a
consequence, the HT effects are effectively absorbed in the
COMPASS PDFs. A crucial point is that the COMPASS analysis does
not include CLAS data, which are entirely in the pre-asymptotic
region, and for which the HT effects are essential. In Fig. 4 the
COMPASS PDFs corresponding to a positive solution for $\Delta G$
are compared with those obtained by LSS'06. As seen from Fig. 4,
except for $(\Delta u + \Delta\bar u)$ the other PDFs differ,
especially those of the strange quarks and gluons.
\begin{figure}[h]
\begin{center}
\begin{tabular}{cc}
\mbox{\epsfig{figure=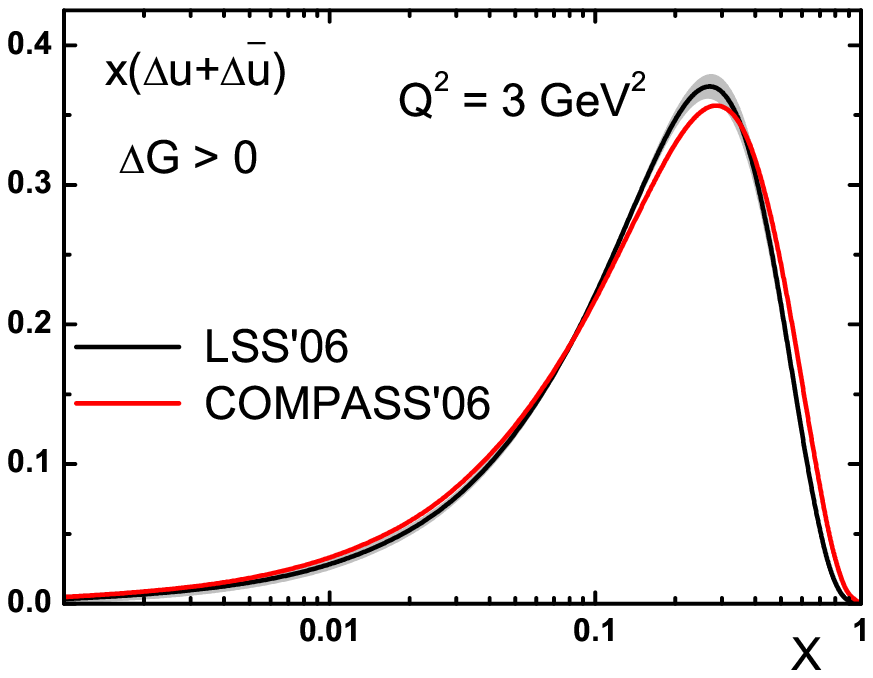,width=5.5cm,height=4.5cm}}&
\mbox{\epsfig{figure=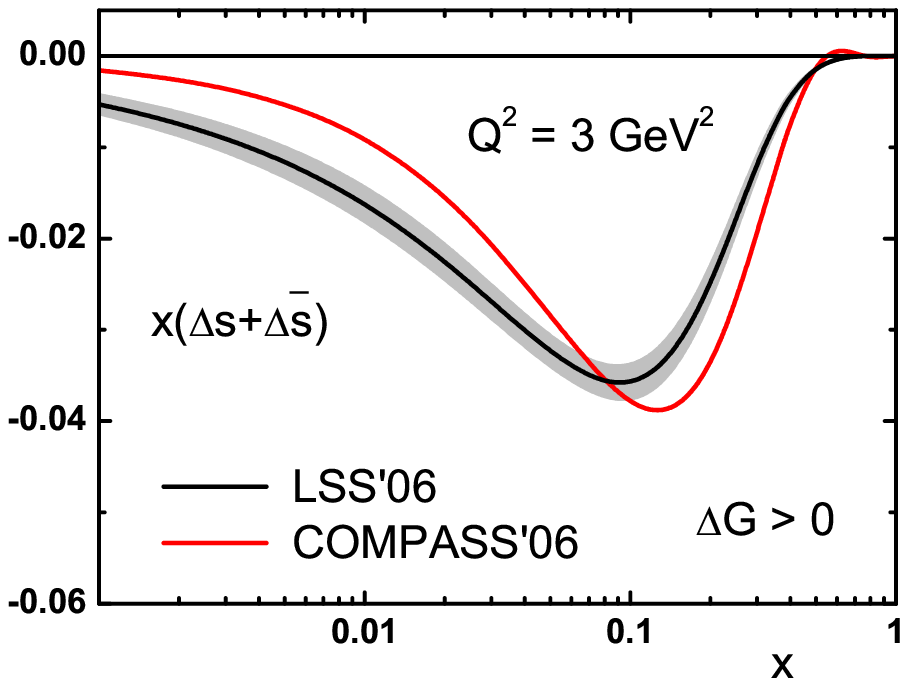,width=5.5cm,height=4.5cm}}\\
\mbox{\epsfig{figure=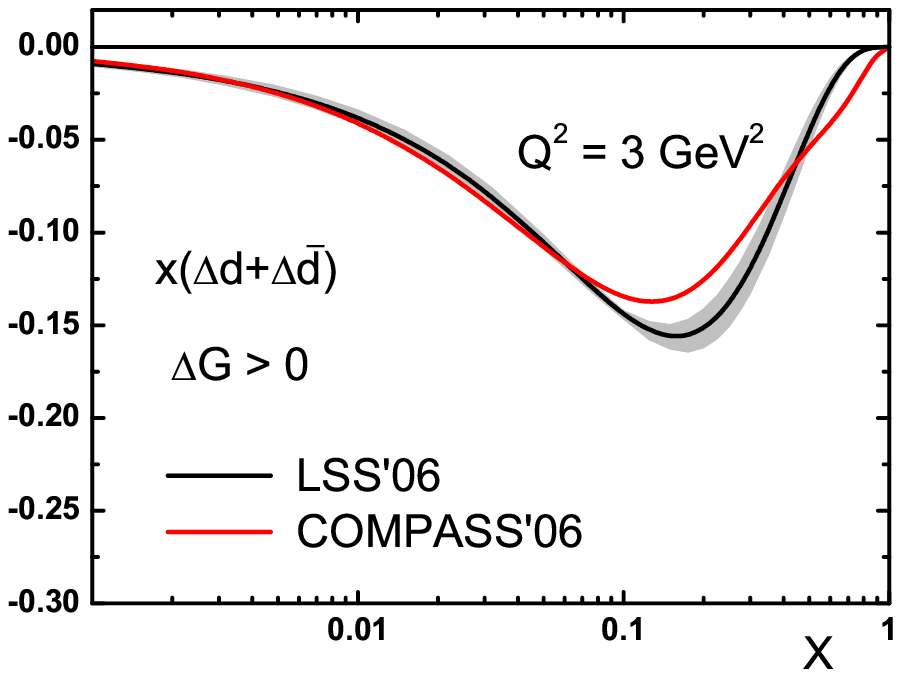,width=5.5cm,height=4.5cm}}&
\mbox{\epsfig{figure=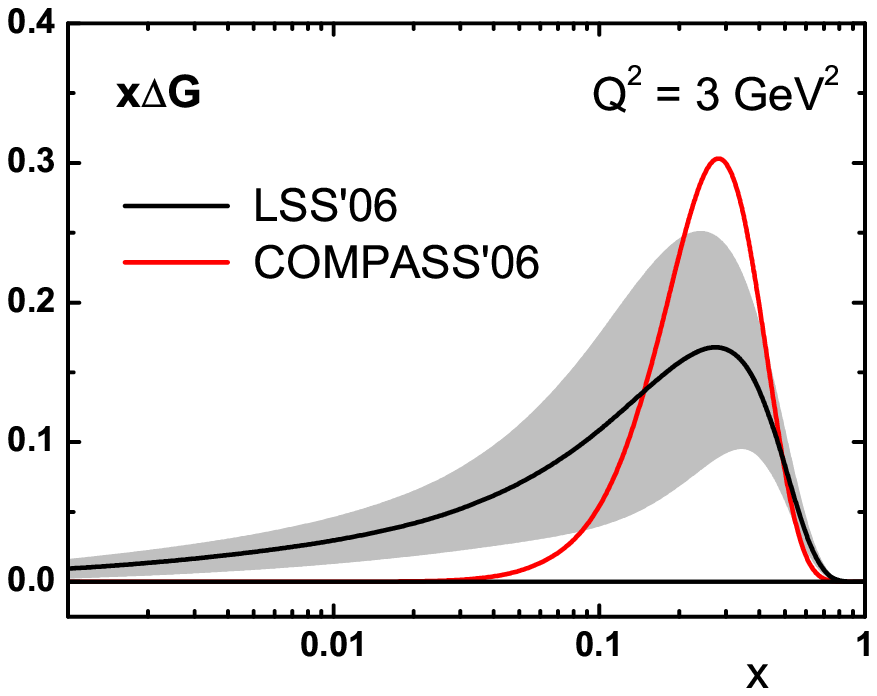,width=5.5cm,height=4.5cm}}\\
\end{tabular}
\end{center}
{\small {\bf Figure 4.} Comparison between NLO($\rm
\overline{MS}$) LSS'06 polarized PDFs and those obtained by\\
COMPASS. }
\end{figure}

\subsection{Comparison between LSS'06 and DSSV PDFs}

Recently the DSSV group has presented results on 
NLO(${\rm \overline{MS}}$) polarized
PDFs \cite{DSSV} obtained from the first global analysis of 
polarized DIS, SIDIS and RHIC polarized pp scattering data. 
Due to the SIDIS data a flavor decomposition of the polarized sea is
achieved. 
For the fit to the inclusive DIS data the second method (\ref{g1f1_LT}) 
was used, i.e., a NLO QCD approximation for $(g_1)_{\rm LT}$ and 
$(F_1)_{\rm LT}$ in the ratio $g_1/F_1$. The unpolarized structure 
function $F_1(x,Q^2)_{\rm LT}$ was calculated using the NLO MRST'02 
parton densities \cite{MRST02}. The difference between 
$F_1(x,Q^2)_{\rm NLO}$ and the phenomenological parametrization of the 
data, $F_1(x,Q^2)_{\rm exp}$, used in our analysis \cite{LSS06} is 
illustrated in Fig. 5. 
It is a measure of the size of the TM and HT 
corrections $F_1(x,Q^2)_{\rm {TMC+HT}}$ to $F_1$ which cannot be ignored 
in the pre-asymptotic region. Note that in the MRST fit to the 
unpolarized data the preasymptotic region was excluded precisely in order to eliminate the TM and HT corrections. 
\begin{figure}[h]
\begin{center}
\begin{tabular}{cc}
\mbox{\epsfig{figure=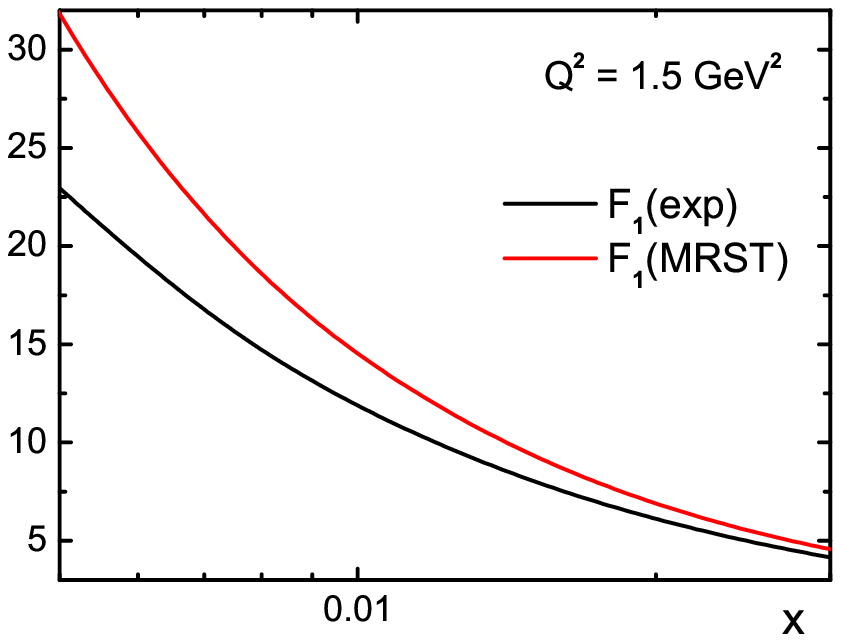,width=5.5cm,height=4.5cm}}&
\mbox{\epsfig{figure=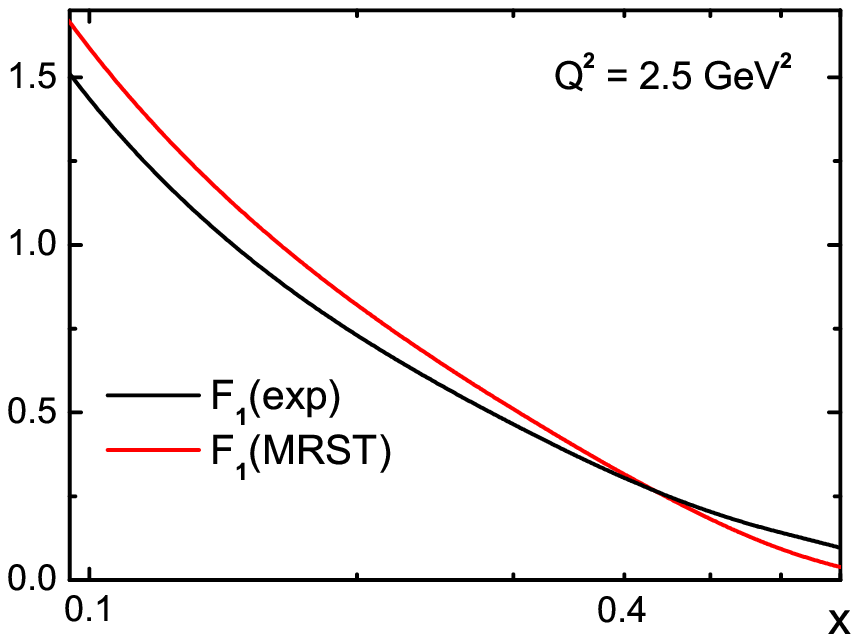,width=5.5cm,height=4.5cm}}\\
\end{tabular}
\end{center}
{\small{\bf Figure 5.} Comparison between $F^p_1(\rm MRST)_{NLO}$ and 
$(F^p_1)_{\rm exp}$ unpolarized structure functions at $Q^2=1.5~GeV^2$ 
(left)  and $Q^2=2.5~GeV^2$ (right).} 
\end{figure}
That is why $F_1(\rm MRST)_{NLO}$ differ from $(F_1)_{\rm exp}$ in the preasymptotic region where the TM 
and HT corrections to $F_1$ cannot be ignored. 

It is important to 
mention also that in the preasymptotic region for large $x$ and lower 
$Q^2$ ($x > 0.40,~0.47$ for JLab and SLAC/E143 data, respectively),
the kinematic factor $\gamma^2=4M^2x^2/Q^2$ is larger than $R(x,Q^2)$. 
Then it follows from the relation between $F_1$ and $F_2$
\begin{equation}
2xF_1(x,Q^2)=F_2(x,Q^2){{1+\gamma^2}\over {1+R(x,Q^2)}}
\label{F1F2}
\end{equation}  
that in this region $2xF_1 > F_2$ and, as a consequence, 
\begin{equation}
F_L(x,Q^2) = F_2(x,Q^2) - 2xF_1(x,Q^2)
\label{FL}
\end{equation} 
is negative, in contrast to what follows from pQCD, i.e. that 
$(F_L)_{\rm LT}$ should be always positive. So, $F_L$ could become 
negative in the preasymptotic region due to HT corrections. We consider 
it is important to test this observation by fitting the data on 
unpolarized structure functions $F_2$ and $F_L$ which will become 
available at JLab in the near future. 

The main features of the results of the fits obtained by LSS (Method I) 
and DSSV (Method II) are illustrated in Fig. 6 for a proton target. 
As expected, the curves corresponding to the ratios $g_1^{tot}(\rm LSS)/(F_1)_{exp}$ and 
$g_1(\rm DSSV)_{NLO}/F_1(\rm MRST)_{NLO}$ practically coincide although different expressions were used for $g_1$ and $F_1$ in the fit 
(see the left panel of Fig. 6; the difference between them for $x>0.2$ 
will be discussed later). 
\begin{figure}[h]
\begin{center}
\begin{tabular}{cc}
\mbox{\epsfig{figure=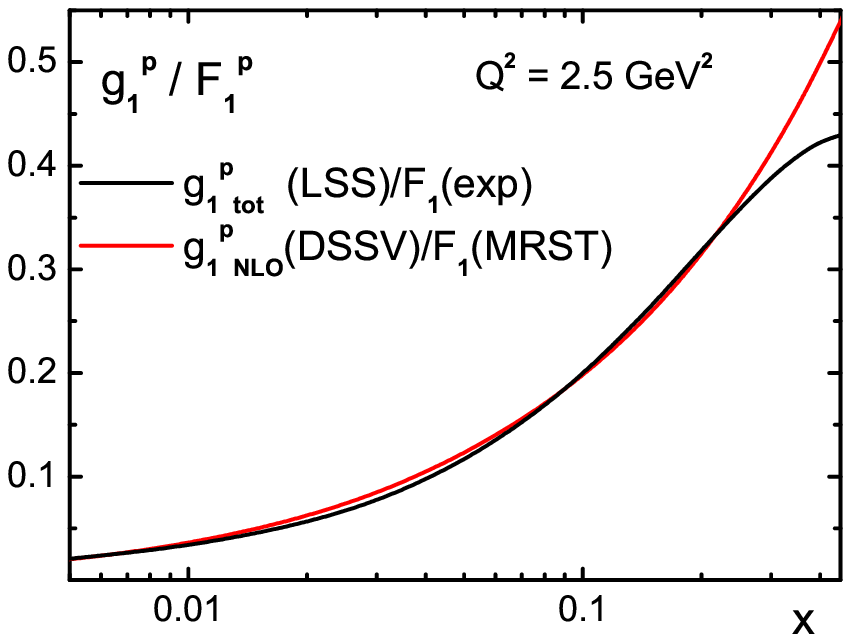,width=6.5cm,height=5cm}}&
\mbox{\epsfig{figure=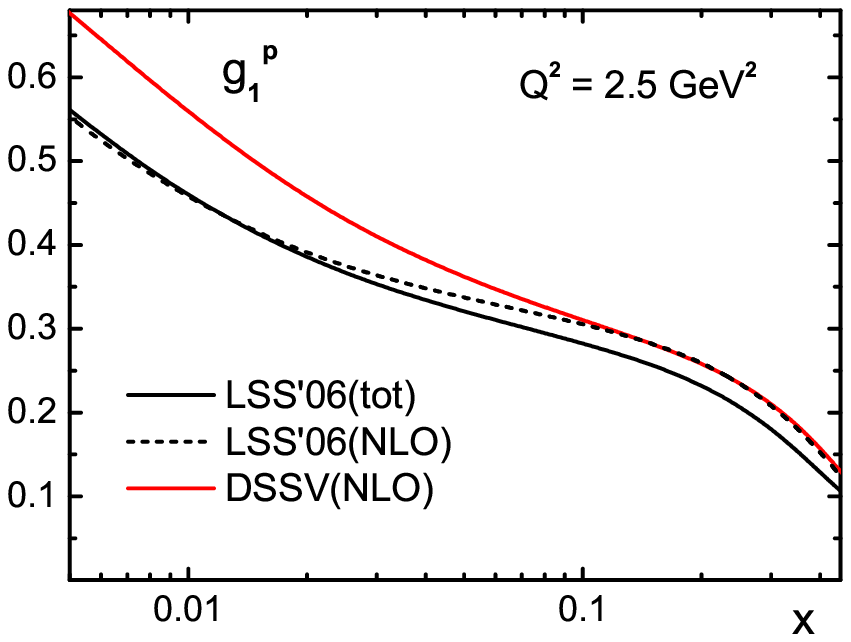,width=6.5cm,height=5cm}}\\
\end{tabular}
\end{center}
{\small{\bf Figure 6.} Comparison between: the ratios $g_1^{tot}(\rm LSS)/(F_1)_{exp}$ and $g_1(\rm DSSV)_{NLO}/F_1(\rm MRST)_{NLO}$ {\bf (left)}; $g_1(\rm LSS)_{NLO}$ and $g_1(\rm DSSV)_{NLO}$ {\bf(right)}. 
The LSS results correspond to the node-type solution for 
$x\Delta G(x,Q^2)$ .} 
\end{figure}
In the right panel of Fig. 6 the LSS and DSSV LT(NLO) pieces of $g_1$ 
are compared for a proton target.
Surprisingly they coincide for $x > 0.1$ although the HT corrections, 
taken into account in LSS'06 and ignored in the DSSV analysis, do NOT 
cancel in the ratio $g_1/F_1$ in this region, as has already been 
discussed above. The understanding of this puzzle is connected with 
the fact that in the DSSV fit to all available $g_1/F_1$ data a factor $(1+\gamma^2)$ was introduced on the RHS of Eq. (\ref{g1f1_LT}) 
\begin{equation}
\left[{g_1(x, Q^2)\over F_1(x, Q^2)}\right]_{exp}~\Leftrightarrow~
{{g_1(x,Q^2)_{\rm LT}}\over {(1+\gamma^2)F_1(x,Q^2)_{\rm LT}}}~. 
\label{g1f1_LT_gamma}
\end{equation}
(Note that for the fit to the $A_1$ data Eq. (\ref{g1f1_LT}) was used.) 

There is no rational explanation for such a correction. The authors 
point out \cite{Sassot} that it is impossible to achieve a good 
description of the $g_1/F_1$ data, especially of the CLAS ones, without 
this correction (see Fig. 7). As seen from Fig. 7, the theoretical 
curves lie systematically above the data which are badly fitted without introducing the $(1+\gamma^2)$ factor. 
\begin{figure}[h]
\begin{center}
\begin{tabular}{cc}
\mbox{\epsfig{figure=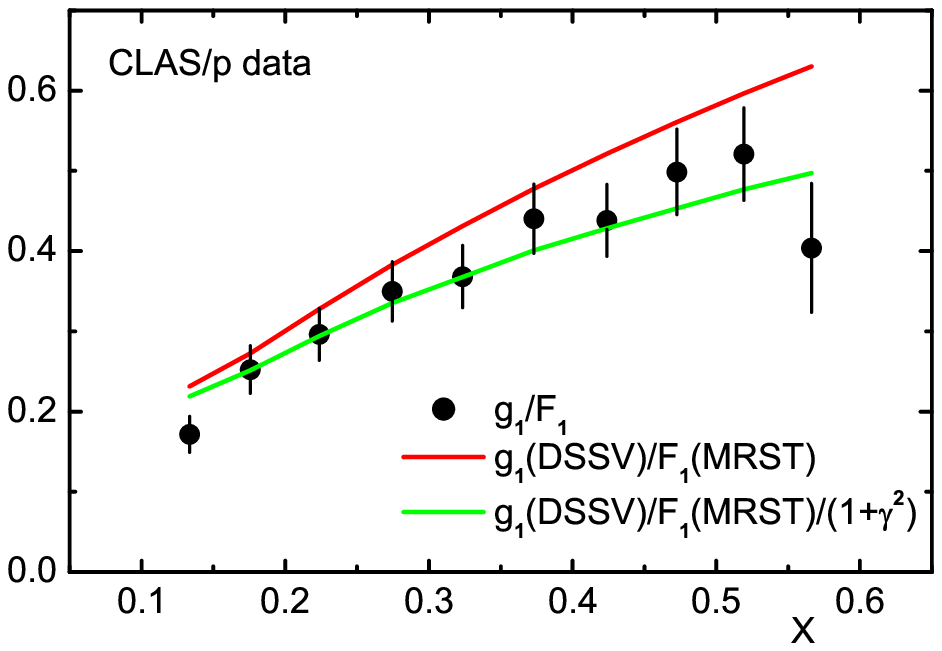,width=6.5cm,height=5cm}}&
\mbox{\epsfig{figure=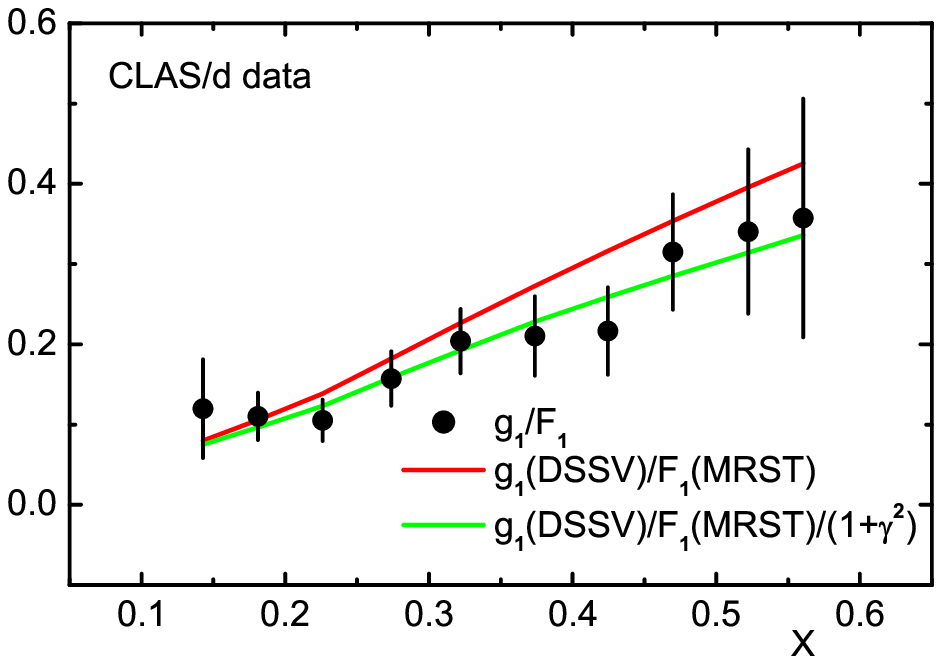,width=6.5cm,height=5.2cm}}\\
\end{tabular}
\end{center}
{\small{\bf Figure 7.}  CLAS $g_1/F_1$ data compared to the 
theoretical DSSV curves accounted or not for the $(1+\gamma^2)$ 
factor.} 
\end{figure}

It turns out empirically that the $1/(1+\gamma^2)$ factor accidentally 
more or less accounts for the TM and HT corrections to $g_1$ and $F_1$ 
in the ratio $g_1/F_1$. The relation
\begin{equation}
 1 + {(g_1)_{\rm TMC+HT}\over (g_1)_{\rm LT}} - {(F_1)_{\rm TMC+HT}
\over (F_1)_{\rm LT}} \approx {1\over {(1+\gamma^2)} }
\label{TMCHT_gamma}
\end{equation}
is satisfied with an accuracy between 4\% and 18\%  for the CLAS 
proton data ($x>0.1$ and $Q^2$ between 1 and 4 $GeV^2$). That is the 
reason why the LT(NLO) pieces of $g_1^p$ obtained by LSS and DSSV 
are in a good agreement for $x>0.1$ (see the right panel of fig. 6). 
Also, why the curve in Fig 6 (left) corresponding to $g_1(\rm DSSV)_{NLO}/F_1(\rm MRST)_{NLO}$ lies above the one of 
$g_1^{tot}(\rm LSS)/(F_1)_{exp}$. Including the $(1+\gamma^2)$ factor in (\ref{g1f1_LT_gamma}) would make the curves almost identical. 
It is important to mention that introducing the $(1+\gamma^2)$ factor 
does not help at $x<0.2$ because it is small in this region and cannot mimic the difference between TM and HT corrections to $g_1$ 
and $F_1$ (LHS of Eq. (\ref{TMCHT_gamma})) for proton as well as for 
neutron target (see Fig. 2). That is why the LT(NLO) pieces of $g_1^p$ obtained by LSS and DSSV groups differ in this region - the smaller $x$ is, the greater is the difference (see Fig. 6 (right)). To summarize: It is impossible to describe the precise data in the preasymptotic region like the CLAS data using the second method. Its 
empirical modification by introducing the $(1+\gamma^2)$ factor accounts approximately for the TM and HT effects, but only in the 
x region: $x>0.1,~0.2$ for proton and neutron targets, respectively.

The NLO($\rm \overline{MS}$) PDFs determined from LSS'06 and DSSV 
analyses are compared in Fig. 8. The AAC'08 PDFs \cite{AAC08} 
obtained from a combined NLO QCD analysis of inclusive DIS and RHIC 
$\pi^0$-production data are also presented. 
\begin{figure}[h]
\begin{center}
\begin{tabular}{cc}
\mbox{\epsfig{figure=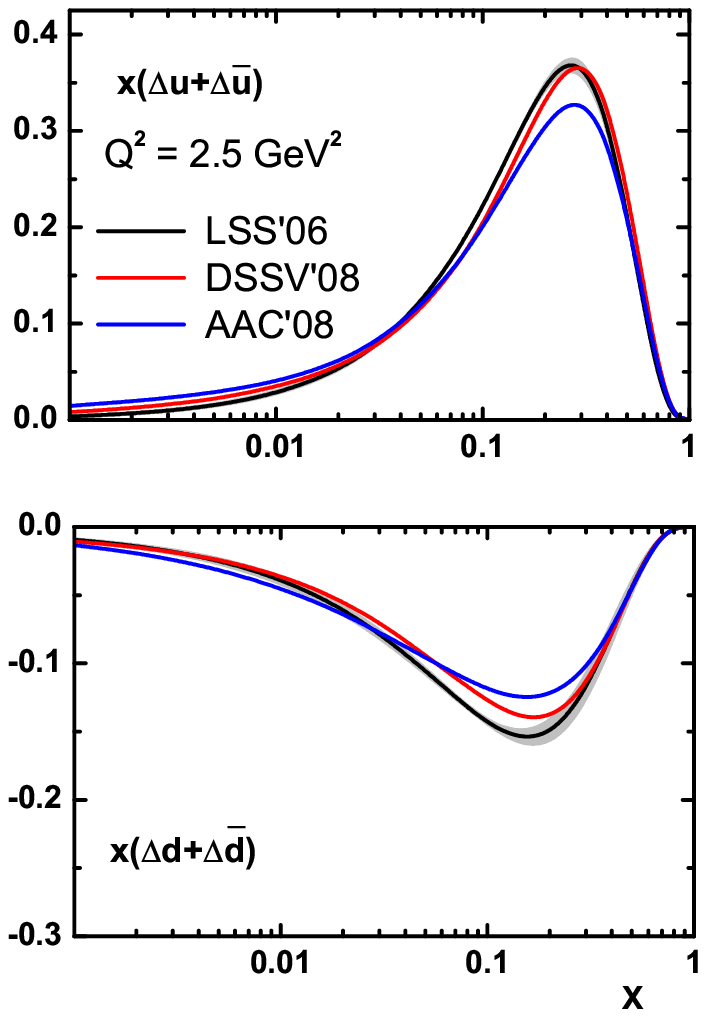,width=5.5cm,height=7.5cm}}&
\mbox{\epsfig{figure=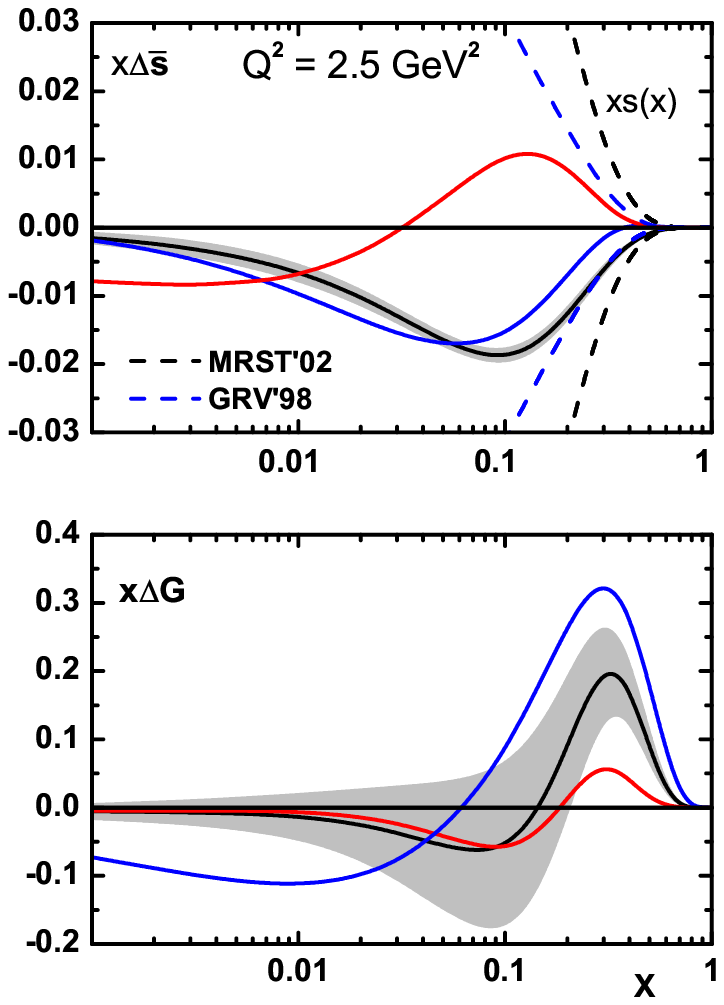,width=5.5cm,height=7.7cm}}\\
\end{tabular}
\end{center}
\centerline{
{\small{\bf Figure 8.} Comparison between LSS'06, DSSV and AAC'08 
NLO PDFs in ($\rm \overline{MS}$) scheme.  } }
\end{figure}
Note that for the fit 
to DIS data AAC have used the modification (\ref{AAC}) of the second 
method and that the RHIC data impact mainly on $\Delta G$.
Note also that all these analyses include the precise CLAS data in the 
preasyptotic region. The results are presented for 
the sums $(\Delta u + \Delta\bar{u})$ and $(\Delta d + \Delta\bar{d})$ because they can only be separated using SIDIS data (the DSSV analysis). 
Although the first moments obtained for the PDFs are almost identical, 
the polarized quark densities themselves are different, especially 
$\Delta \bar{s}(x)$ (in all the analyses $\Delta s(x)=\Delta \bar{s}(x)$ is assumed).

Let us discuss the impact of HT effects on 
$(\Delta u + \Delta\bar{u})$ and $(\Delta d + \Delta\bar{d})$ parton densities which should be well determined from the inclusive DIS data.   
$(\Delta u + \Delta\bar{u})$ extracted by LSS and DSSV are well 
consistent. As was discussed above the HT effects for the proton 
target are effectively
accounted for in the DSSV analysis for $x>0.1$ by the introduction of the $(1+\gamma^2)$ factor. However, this factor cannot account for the HT effects for a neutron target at $x<0.2$ (see Fig. 2) and the impact 
of higher twist on $(\Delta d + \Delta\bar{d})$ determined by DSSV is demonstrated in Fig. 8. The positive HT effects are absorbed into 
$(\Delta d + \Delta\bar{d})_{\rm DSSV}$ and it is thus less negative in 
this region. The influence of HT effects at small x (not accounted 
for by the DSSV group) on both parton densities is not sizable because 
of two reasons: first, their values are small and second, the data in 
this region are not precise enough to indicate the impact of higher 
twist on their values. The impact of HT effects on both $(\Delta u + \Delta\bar{u})_{\rm AAC}$ and $(\Delta d + \Delta\bar{d})_{\rm AAC}$ 
is larger because the AAC Collaboration has not taken them into account 
 at all, and in addition, the incorrect approximation $A_1 \approx g_1/F_1$ for some of the data in the preasymptotic region has been used. 
The difference between the strange sea densities $\Delta\bar{s}(x,Q^2)_{\rm LSS}$ and $\Delta\bar{s}(x,Q^2)_{\rm AAC}$ for $x>0.1$ is due to the different positivity conditions which have been used by the two groups. 
Note also that the positivity condition $\vert\Delta G(x,Q^2)\vert \leq G(x,Q^2)$ is not satisfied for the polarized gluon density obtained by 
AAC, which suggests it is not physical. 

In contrast to a negative $\Delta \bar{s}(x,Q^2)$ obtained in all 
analyses of inclusive DIS data, the DSSV global analysis yields a changing in sign $\Delta \bar{s}(x,Q^2)$: positive for $x > 0.03$ 
and negative for small x. 
Its first moment is negative (practically fixed by the SU(3)
symmetric value of $a_8$) and almost identical with that obtained
in the inclusive DIS analyses. It was shown \cite{delsCOMPASS} that 
the determination of $\Delta \bar{s}(x)$ from 
SIDIS strongly depends on the fragmentation functions (FFs) and the new FFs \cite{FF} are crucially responsible for the unexpected behavior of 
$\Delta \bar{s}(x)$. So, obtaining a final and unequivical result for 
$\Delta \bar{s}(x)$ remains a challenge for further research on the internal spin structure of the nucleon.
 
\section{Summary}

The fact that more than a half of the present polarized DIS data are 
in the preasymptotic region makes the QCD analysis of the data more 
complex and difficult. In contrast to the unpolarized case, the 
$1/Q^2$ terms ({\it kinematic} - $\gamma^2$ factor, target mass 
corrections, and {\it dynamic} - higher twist corrections to the 
spin structure function $g_1$) cannot be ignored, and their role in determining the polarized PDFs is important. Sets of polarized PDFs extracted from the data using different methods of QCD analysis, 
accounting or not accounting for the kinematic and dynamic $1/Q^2$ corrections, are considered. The impact of higher twist effects on the determination of the parton densities is demonstrated. It is pointed out that the very accurate DIS data in the preasymptotic region require a more careful matching of QCD to the data in order to extract the polarized PDFs correctly.
 
\vskip 1.5cm
\begin{center}
{\bf Acknowledgments}
\end{center}

This research was supported by the JINR-Bulgaria Collaborative
Grant, by the RFBR Grants (No 08-01-00686, 09-02-01149) and by the
Bulgarian National Science Foundation under Contract 288/2008.
One of the authors (D.S) is grateful to Theory Division at CERN for 
providing the facilities essential for the completion of this work.



\begin{thebibliography}{9}

\bibitem{g1F1data}
K.V. Dharmawardane {\it et al.} (CLAS Collaboration), Phys. Lett. B
{\bf 641}, 11 (2006); X. Zheng {\it et al.} (JLab/Hall A
Collaboration), Phys. Rev. Lett. {\bf 92}, 012004 (2004); K. Abe
{\it et al.} (SLAC E143 Collaboration), Phys. Rev. D {\bf 58},
112003 (1998); P.L. Anthony {\it et al.} (SLAC E155
Collaboration), Phys. Lett. B {\bf 463}, 339 (1999); {\bf 493}, 19
(2000).

\bibitem{CERNdata}
J. Ashman {\it et al.} (EMC Collaboration), Phys. Lett. B {\bf
206}, 364 (1988); Nucl. Phys. {\bf B328}, 1 (1989); B. Adeva {\it
et al.} (SMC Collaboration) Phys. Rev. D {\bf 58}, 112001 (1998).

\bibitem{COMPASS}
V.Y. Alexakhin {\it et al.} (COMPASS Collaboration), Phys. Lett. B
{\bf 647}, 8 (2007).

\bibitem{HERMES}
A. Airapetian et al. (HERMES Collaboration), Phys. Rev. {\bf D71},
012003 (2005).

\bibitem{SLAC_A1}
P.L. Anthony et al. (SLAC E142 Collaboration), Phys. Rev. {\bf
D54}, 6620 (1996); K. Abe et al. (SLAC/E154 Collaboration), Phys.
Rev. Lett. {\bf 79}, 26 (1997).

\bibitem{HERMES07}
A. Airapetian et al. (HERMES Collaboration), Phys. Rev. {\bf D75},
012007 (2007).

\bibitem{LSS_HT}
E. Leader, A.V. Sidorov, and D.B. Stamenov, Phys. Rev. D {\bf 67},
074017 (2003).

\bibitem{GRSV}
M. Gl\"{u}ck, E. Reya, M. Stratmann, and W. Vogelsang, Phys. Rev.
D {\bf 63}, 094005 (2001).

\bibitem{TMC}
A.V. Sidorov and D.B. Stamenov, Mod. Phys. Lett. {\bf A21}, 1991 
(2006).

\bibitem{LSS06} E. Leader, A.V. Sidorov, and D.B. Stamenov,
Phys. Rev. D {\bf 75}, 074027 (2007).

\bibitem{Alekhin}
S.I. Alekhin, Phys. Rev. D {\bf 68}, 014002 (2003).

\bibitem{HTcancel}
E. Leader, A.V. Sidorov, and D.B. Stamenov, in the Proceedings of 
16th International Workshop on
Deep Inelastic Scattering and Related Subjects (DIS2008), 7-11
April, 2008, London, UK (edited by R. Devenish and J. Ferrando,
Science Wise Publishing, 2008, 206 (arXiv:0806.2094 [hep-ph]).

\bibitem{Stamenov}
D.B. Stamenov, unpublished. 

\bibitem{BB}
J. Blumlein, H. Bottcher, Nucl. Phys. {\bf B 636}, 225 (2002).

\bibitem{AAC}
AAC, M. Hirai et al., Phys. Rev. {\bf D 69}, 054021 (2004).

\bibitem{DSSV}
D. de Florian, R. Sassot, M. Stratmann, and W. Vogelsang, Phys.
Rev. Lett. {\bf 101}, 072001 (2008); arXiv:0904.3821 [hep-ph].

\bibitem{AAC08}
M. Hirai and S. Kumano, arXiv:0808.0413 [hep-ph].

\bibitem{Winmolders}
R. Winmolders, private communication.

\bibitem{MRST02}
A.D. Martin, R.G. Roberts, W.J. Stirling, and R.S. Thorne, Eur.
Phys. J. C {\bf  28}, 455 (2003).

\bibitem{Sassot}
R. Sassot, private communication.

\bibitem{delsCOMPASS}
M. Alekseev et. al. (COMPASS Colaboration), arXiv:0905.2828 [hep-ex].

\bibitem{FF}
D. de Florian, R. Sassot, and M. Stratmann, Phys. Rev. D {\bf 75},
114010 (2007); D {\bf 76}, 074033 (207).


\end{thebibliography}
\end{document}